\documentclass[runningheads]{llncs}
\usepackage{graphicx}
\usepackage{subfig}
\usepackage{float}
\usepackage{verbatim}
\begin{document}
\title{Exercise Hierarchical Feature Enhanced Knowledge Tracing}
%
\author{Hanshuang Tong\inst{1,2}\and
	Yun Zhou\inst{1,3} \and
	Zhen Wang\inst{1,4}}
\authorrunning{}
%

\institute{AIXUEXI Education Group Ltd, AI Lab \and
	\email{tonghanshuang2018@163.com} \and
	\email{zhouyun.nudt@gmail.com} \and
	\email{WangZhen\_\_00@163.com}}

\maketitle              
\begin{abstract}
Knowledge tracing is a fundamental task in the computer-aid educational system. 
In this paper, we propose a hierarchical exercise feature enhanced knowledge 
tracing framework, which could enhance the ability of knowledge tracing by 
incorporating knowledge distribution, semantic features, and difficulty  
features from exercise text. Extensive experiments show 
the high performance of our 
framework.

\end{abstract}
\section{Introduction}
Knowledge tracing is an essential and classical problem in intelligent 
education systems. By tracing the knowledge transition process, we could 
recommend specific educational items to a student based on one's weak 
knowledge. Existing methods try to solve knowledge tracing problems from both 
educational psychology and data mining perspectives, such as Item Response 
Theory (IRT) 
\cite{IRT}, Bayesian Knowledge Tracing (BKT)~\cite{BKT}, Performance 
Factors Analysis (PFA) framework~\cite{PFA} and Deep knowledge tracing 
(DKT)~\cite{DKT}. Those models have been proved effective but still have 
limitations. They do 
not systematically consider the impact of different attributes of the exercises 
itself on the knowledge tracing problem. Exercise Enhanced Knowledge 
Tracing(EKT)~\cite{EKT} is the first method to take exercise text and 
attention mechanism 
into consideration. However, EKT extracts features of text by feeding the text 
of exercise directly into a neural network, which fails to extract hierarchical 
features from exercise.    

\begin{figure}\centering
\includegraphics[width=8.3cm]{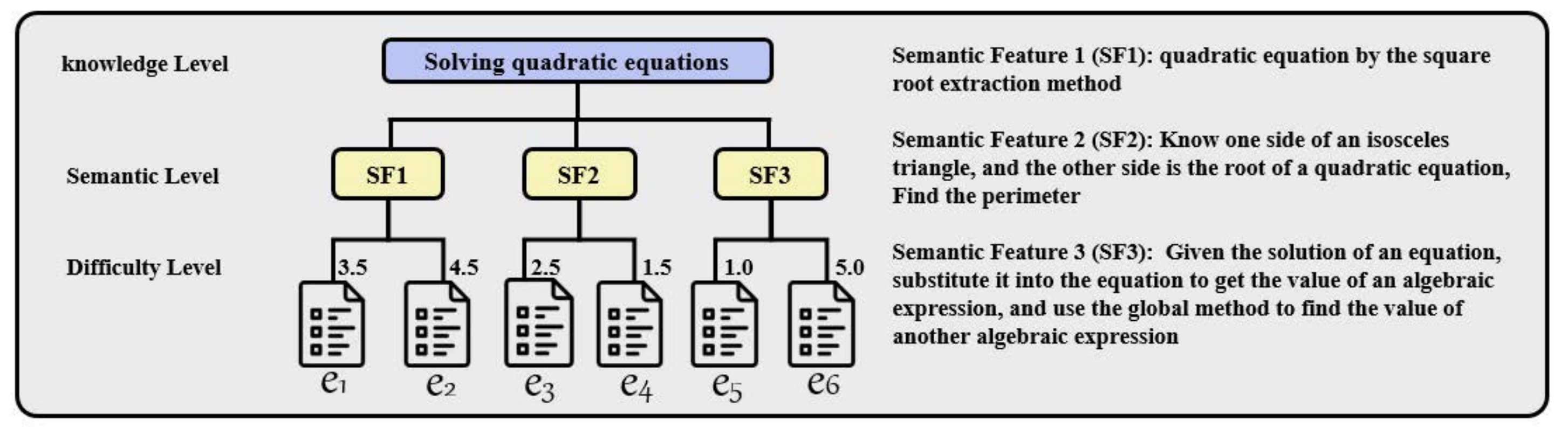}
\caption{The illustration of hierarchical features of exercise} \label{level}
\end{figure}

\begin{figure}[htb]\centering
	\includegraphics[width=8cm]{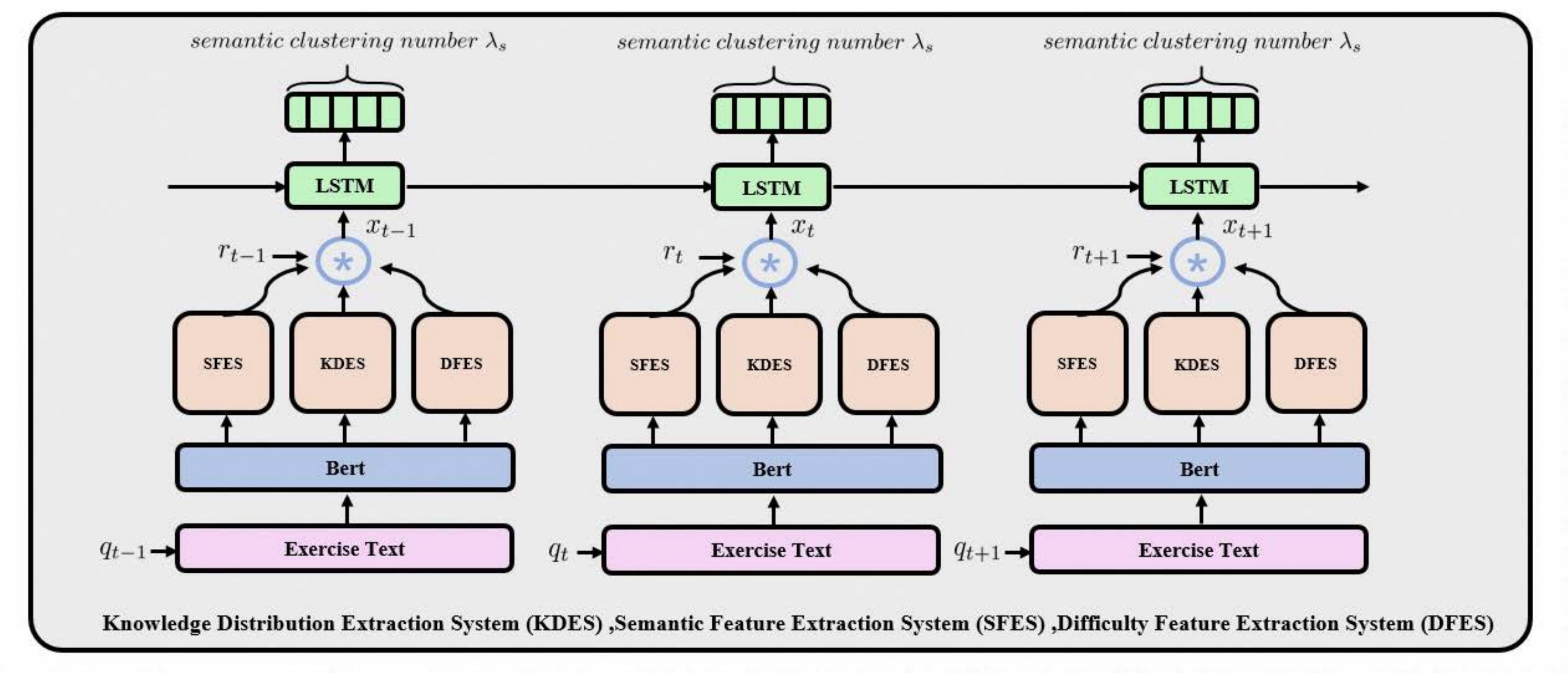}
	\caption{Exercise hierarchical feature enhanced framework.} \label{frame}
\end{figure}  

\section{Exercise Hierarchical Feature Enhanced Framework}

\subsubsection{Framework Overview}
Knowledge tracing task can be summarized as : In an online educational system,  
suppose we have $M$ students and $E$ exercises in total. Given any learners’ 
exercise record $ E =\{(q_{1},r_{1}),(q_{2},r_{2})…(q_{m},r_{m})\}$, predict 
one's performance on $q_{t+1}$. Here $(q_{t},r_{t})$ represents that a 
learner practices question $q_{t}$ and answers $r_{t}$ at step t. The entire 
structure of the framework 
is shown in Fig.~\ref{frame}. In order to dig deeper into the information in 
the exercise text, first we utilize Bert~\cite{Bert} to generate embedding 
vector $v_{b}$. Then we feed them into three systems to generate 
knowledge distribution $v_{t} \in R^{K}$, semantic features $s_{t}$ 
and question difficulty $d_{t}$ separately. Let $\varphi(s_{t})$ be the one-hot 
encoding of the semantic cluster where the question belongs at time t. Finally, 
we concatenate $v_{t}$, $\varphi(s_{t})$, $d_{t}$, and $r_{t}$ as $x_{t}$ and 
feed $x_{t}$ into a sequence model.

\subsubsection{Subsystems Introduction}

Two text classification systems, named KDES and DFES, are designed to predict 
the knowledge distribution and difficulty of the exercise respectively. The 
semantic feature extractor system(SFES) could be considered as an unsuperviesed 
clusering problems. The input of those systems is 
the Bert encoding of the exercise text. The knowledge labeled by teacher and 
the correct rate of a question~\cite{hontangas2000choice} serve as ground 
truth and are predicted 
using 
TextCNN~\cite{TextCNN} in KDES and DFES systems. In KDES system, we use 
softmax 
results classified in the trained model to represent the knowledge distribution 
of an exercise. In DFES systems, we use neural networks to predict difficulty 
in order to solve the cold start problem. In SFES systems, 
we cluster the input using a Hierarchical Clustering method by calculating the 
cos distance between different semantic vectors~\cite{johnson1967hierarchical}.

 \begin{equation}
 h_{t},c_{t} = LSTM(x_{t},h_{t-1},c_{t-1};\theta_{t})\label{con:e1}
 \end{equation}  
 
 \begin{equation}
 y_{t}= \sigma(W_{yh} \cdot h_{t}+b_{y})\label{con:e2}
 \end{equation}
 
 \begin{equation}
 loss = 
 -\sum_{t}(r_{t+1}*log(y_{i}^{T}\cdot{\varphi(s_{t+1})})+(1-r_{t+1})
 *log(1-y_{i}^{T}\cdot{\varphi(s_{t+1})})\label{con:e3})
 \end{equation}

 \subsubsection{Modeling Process}
 In the propagation stage, as shown in Equation \ref{con:e1}, we process 
 $x_{t}$ and the 
 previous learner's hidden state $h_{t-1}$ and then use RNN network to get 
 current 
 learner's hidden state $h_{t}$. Here we use LSTM as a variant of RNN since it 
 can better preserve long-term dependency in the exercise 
 sequence~\cite{memeory}. Finally, we use $h_{t}$ to predict $y_{t}$ which 
 contains information about students' mastery of each semantic feature. 
 Additionally, the dimension of $y_{t}$ is same as the total number of 
 different semantic clustering in DFES system. The $\theta_{t}, W_{yh}, b_{y}$ 
 in the equation 
 are the parameters of models. The goal of training is to minimize the negative 
 log 
 likelihood of the observed sequence of student response logs(shown in 
 Equation \ref{con:e3}).

\section{Experiment}

\subsection{Experimental Setting}
Since there is no open dataset which could provide exercising records with text 
information. We derive an experimental dataset containing 132,179 students and 
91,449,914 answer records from a large real-world online education system : 
aixuexi.com.

The baselines of the experiments are as following: BKT, which is based on 
Bayesian 
inference; DKT, which uses recurrent neural networks to model student learning; 
EKTA, which incoporate exercise text features and attention mechanism into 
the recurrent neural networks; EHFKT\textunderscore K/S/D, a simplified version 
of EHFKT, which only contains KDES/SFES/DFES system. The input of EHFKT series 
is the concatenation of problem encoding and the ouput of each system; 
EHFKT\textunderscore T, 
which contains all subsystems. It diagnoses the transition of mastery of 
knowledge, while EHFKT diagnoses transition of the mastery of semantic 
features.

\subsection{Experimental Results}

\subsubsection{Hierarchical Clustering Result}

\begin{figure}
\begin{minipage}[t]{0.5\linewidth}
\centering
\includegraphics[width=6cm]{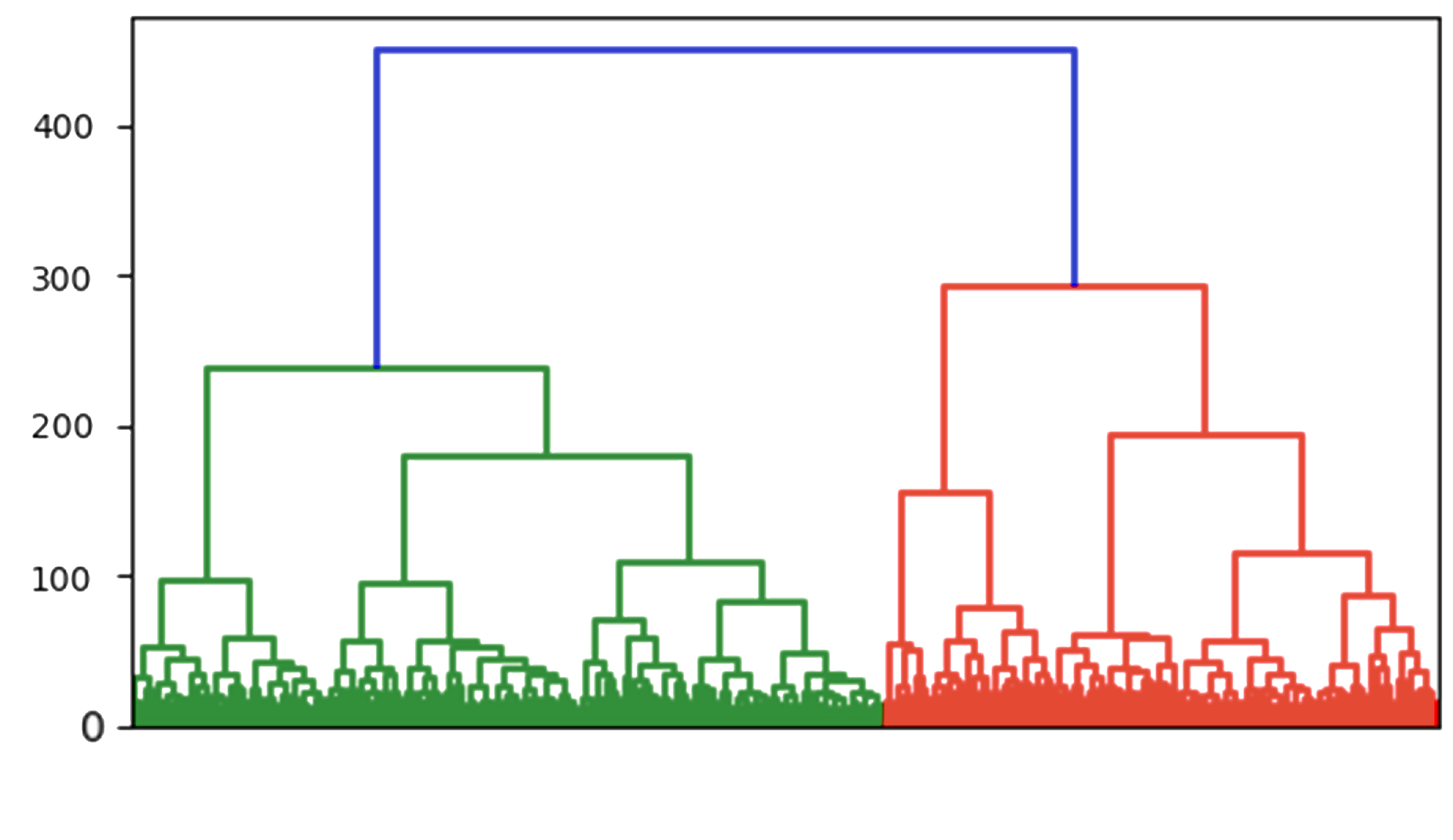}
\caption{Hierarchical clustering result}
\label{sfes_res}
\end{minipage}%
\begin{minipage}[t]{0.5\linewidth}
\centering
\includegraphics[width=4cm]{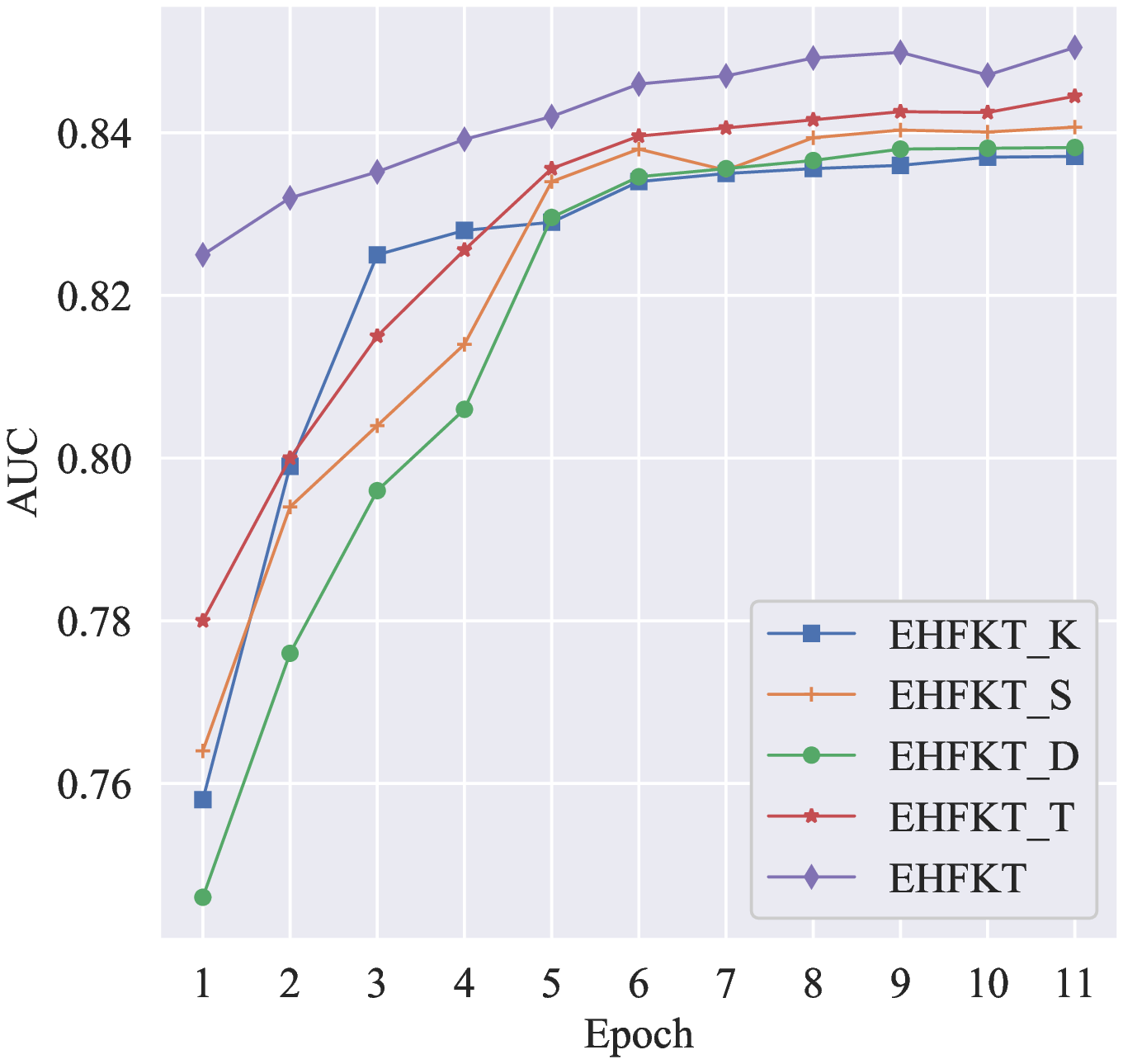}
\caption{AUC of EHFKT series}
\label{ehfkt_auc}
\end{minipage}
\end{figure}

The SFES system uses Bert and Hierarchical Clustering to obtain semantic 
features of questions. Fig.~\ref{sfes_res} shows the visualization of the 
clustering results of 11410 questions. The y-axis corresponds to the 
classification threshold and x-axis corresponds to each exercise. 
Table~\ref{semantic} implies the result of clustering when the number of 
clustering $\lambda_{s}$ is 912.

\begin{table}
\caption{The result of clustering}\label{semantic}
\begin{tabular}{|p{0.6cm}|p{8.4cm}|p{2cm}|p{1cm}|}
\hline
Id & Question Content & Knowledge & Cluster\\
\hline
{Q35} &{Calculate factorization of 
$9{{a}^{2}}\left(2x-3y\right)+4{{b}^{2}}\left(3y-2x\right)$} & {Factorization} 
& {SF3} \\
\hline
{Q37} &{Calculate $\left(2{{a}^{3}}+{{a}^{2}}\right)\div{{a}^{2}}$,which of the 
following is true?} & {Factorization} & {SF3} \\
\hline
{Q38} &{If 
$\left({{x}^{2}}+{{y}^{2}}+2\right)\left({{x}^{2}}+{{y}^{2}}-2\right)=0$, 
calculate ${{x}^{2}}+{{y}^{2}}$} & {Factorization} & {SF4} \\
\hline
{Q36} &{Given two points on $y = - m {{x} ^ {2}} + 2 x $, calculate m} & 
{Factorization} & {SF5} \\
\hline
\end{tabular}
\end{table}

\subsubsection{EHFKT Result}
In this part, our experiment divides the dataset into a training set with 
105,744 learners' logs and a test dataset with 26,435 learners' logs. 
Fig.~\ref{ehfkt_auc} 
shows the transition of AUC during the training process. Table~\ref{all_auc} 
shows the overall comparing results in this task. The results indicate that 
EHFKT performs better than other baseline models. Thus, we could draw several 
conclusions from the result: In the knowledge tracing task, adding hierarchical 
features can better represent questions; Besides, tracing the mastery of 
semantic clusterings
can predict students' 
performance more precisely. The reason is that the exercises contained in the 
same 
clusters have 
similar knowledge distribution, difficulty, and semantics; This result
also demonstrates the instability of the tracing of knowledge mastery since the 
difficulty of an exercise is unpredictable.

\section{Conclusions}

In this article, we propose a novel knowledge tracing framework which could 
extract the knowledge distribution, semantic features and difficulty from the 
exercise. Besides, We introduce the diagnosis of semantic features of questions 
into knowledge tracing, which leads to more accurate performance prediction. 
Although the meaning of these semantic clusters is beyond people's 
understanding, in the future we will try extracting the meaning of the 
exercises in the 
same cluster by text sumarization technique to make the data-driven clusters 
result more understandable to human.
\begin{table}\centering
	\caption{Evaluation metrics of different deep learning 
		methods}\label{all_auc}
	\begin{tabular}{|p{2.8cm}|p{3cm}|p{2.8cm}|p{3cm}|}
		\hline
		Model  & AUC & Model & AUC  \\
		\hline
		BKT & {$0.6325 \pm 0.0011$}  & DKT & {$0.8324 \pm 0.0031$}  \\
		\hline
		EKTA  & {$0.8384  \pm 0.0036$} &  EHFKT\textunderscore S  & {$0.8407  
		\pm 0.0016$} \\
		\hline
		EHFKT\textunderscore K  & {$0.8371 \pm  0.0022$} &  
		EHFKT\textunderscore D &{$0.8382 \pm 0.0035$}  \\
		\hline
		EHFKT\textunderscore T & {$0.8445  \pm 0.0025$} &  EHFKT &  
		{$\textbf{0.8505} \pm 
			\textbf{0.0021}$} \\
		
		\hline
	\end{tabular}
\end{table}

%
%
%
%

\bibliographystyle{splncs04}
\bibliography{knowledge_tracing}
\end{document}